\documentclass[aps,prd,twocolumn,showpacs,groupedaddress,nofootinbib]{revtex4}
 
\usepackage{graphicx}
\usepackage{longtable}

\newcommand {\Dtau}{\Delta\tau}

\begin{document}

\title{\bf
Renormalization of Anisotropy and Glueball Masses on Tadpole 
Improved Lattice Gauge Action}

\author{Mushtaq Loan}
\email{mushe@phys.unsw.edu.au}
\author{Tim Byrnes}
\author{Chris Hamer}

\affiliation{School of Physics, The University of New South Wales,
Sydney, NSW 2052, Australia}  

\date{January 10, 2003}


\begin{abstract}
The Numerical calculations for tadpole-improved $U(1)$ lattice gauge theory  
in three-dimensions on anisotropic lattices have been performed using 
standard path integral Monte Carlo techniques.
Using average plaquette tadpole renormalization scheme, simulations were 
done with temporal lattice spacings much smaller than the spatial ones and 
results were obtained for the string tension,  
the renormalized 
anisotropy  and scalar glueball masses. We find, by comparing the 
`regular' and `sideways' potentials,  that tadpole improvement results in 
very little renormalization of the bare anisotropy and reduces the 
discretization errors in the static quark potential and in the glueball 
masses. 
\end{abstract}

\maketitle

\section{Introduction}                             
Compact $U(1)$ gauge theory in (2+1) dimensions
is one of the simplest models with dynamical gauge degrees of freedom
and possesses some important similarities with QCD \cite{ham94}.
 The model has two essential features in common with QCD, confinement
\cite{pol77,pol78} and chiral symmetry breaking \cite{fie90}.
 The theory is
interesting in its own right, for it has analytically
been shown to
confine electrically charged particles even in the weak-coupling regime 
(at
zero temperature)\cite{pol77,pol78,ban77,banks78,dre79,gof82}.
The confinement is understood as a result of the dynamics of the
monopoles which emerge due to the compactness of the gauge field. The 
string tension as a function of the coupling behaves in a similar fashion 
to that of the 4-dimensional $SU(N)$ lattice gauge theory. This model 
also  allows us to work with large lattices with reasonable statistics. 
Other common features of
compact $U(1)_{(2+1)}$ and QCD are the existence of a mass gap and of a
confinement-deconfinement phase transition at some non-zero temperature.
Thus, being reasonably simple and theoretically well understood in the
weak-coupling limit, the $U(1)$ model
provides a good testing ground for the development of  new methods and new
algorithmic approaches.
In a recent paper \cite{loan02} we have obtained the first clear picture 
of the 
static quark potential, showing very clear evidence of the linear 
confining behaviour at large distances. The  evidence of the the 
scaling behaviour of the string tension and the mass gap has also been 
observed in this model. 

In the present paper we want to extend the 
analysis of Ref. \cite{loan02} in various respects. Since  the measured
ratio of spatial to temporal lattice spacings is not the same as the input
parameter in the action, it becomes important 
to determine  the true or renormalized anisotropy $\xi_{phys.}$ as a 
function of the bare anisotropy $\xi_{0}$. 
An important advantage of using anisotropic lattices has been the need to 
measure the renormalization of anisotropy in the simulation. 
The existing theoretical \cite{snipp97,garcia97} 
and numerical
studies \cite{klas98} with Wilson action for $SU(3)$ lattice gauge theory
have shown that
at finite coupling $g$, the renormalized anisotropy $\xi_{phys.}$ differs 
appreciably form 
bare anisotropy $\xi_{0} $. However, it has recently been shown that the use 
of improved actions, supplemented by tadpole improvement, besides providing
a better discretization scheme for QCD, offer the advantage of a significant
reduction of renormalization of $\xi_{0} $ to a few percent 
\cite{morn97,mor96}. 
The anisotropy 
parameter $\eta$ (ratio of the renormalized and bare anisotropy) and 
anisotropic
coefficients have been calculated to one-loop order for improved actions in
various recent studies 
\cite{garcia97,sakai00,eng00,alford01,drummond02}. These calculations have
provided very reliable results and the observed behaviour 
is confirmed non-perturbatively by large scale simulations on fine 
lattices \cite{morn97, shakes98,alford98}. We investigate the 
influence of tadpole improvement on isotropic and anisotropic 
lattices for the $U(1)$ model in (2+1) dimensions in 
reducing the renormalization of the bare anisotropy at  weak and  
strong couplings. 
We also apply tadpole improved $U(1)$ lattice gauge theory to 
calculations of the static quark potential, the string tension and the 
scalar glueball masses and compare the results with simulations of the 
Wilson action.    

The rest of this paper is organized as follows. After outlining the 
tadpole improved $U(1)$ gauge model in (2+1) dimensions in Sect. II,  we
describe the method for determination of the renormalized anisotropy in 
Sect. III. We present  results from our simulations on anisotropic 
lattices using both standard and tadpole improved  Wilson gauge action in
Sect. IV. We compare bare and renormalized anisotropies, static quark 
potential and scalar glueball masses from these actions. We conclude in 
Sect. V with a summary and outlook on future work.   
\section{Compact $U(1)$ model in (2+1) dimensions}
The tadpole-improved $U(1)$ gauge action on an anisotropic lattice can be 
written in the following form \cite{klas98}:
\begin{equation}
S = \beta \left[\sum_{r,i>j}\frac{\xi_{0}}{u_{s}^{4}}(1-P_{ij}(r))
+\sum_{r,i}\frac{1}{\xi_{0} u_{s}^{2}u_{t}^{2}}(1-P_{it}(r))\right]
\label{eqn1}  
\end{equation}
where $P_{\mu\nu}$ is the plaquette operator, $\xi_{0} = \Dtau 
=a_{t}/a_{s}$ is the bare anisotropy at the classical level and  $u_{s}$ 
and
$u_{t}$ are the mean fields for the tadpole improvement. 
The notation used in Eq.(\ref{eqn1}) differ slightly from that used in 
Refs. \cite{garcia97,drummond02}, where the spatial and temporal mean-field 
improvement factors, $u_{s}$ and $u_{t}$ were absorbed into 
definition  of $\beta$ 
and $\xi_{0}$. This, however, follows the notation introduced in Ref. 
\cite{morn97}.

On the anisotropic lattice, the mean fields are determined  using the 
measured values of the average plaquettes \cite{mor97}. We first compute 
$u_{s}$ from 
spatial plaquettes, $u_{s}^{4}=\langle P_{ij}\rangle$, and then we compute 
$u_{t}$ from 
temporal plaquettes, $u_{t}^{2}u_{s}^{2}=\langle P_{it}\rangle$. Another way to 
determine 
the mean fields is to use the mean links in Landau gauge \cite{lep98}
\begin{equation}
u_{t} = \langle\mbox{ReTr}U_{t}\rangle, \hspace{1.0cm} u_{s} = \langle\mbox{ReTr}U_{i}\rangle
\label{eqn.2}
\end{equation}
where the lattice version of the gauge condition is obtained by maximizing 
the quantity,
\begin{equation}
\sum_{r,\mu}\frac{1}{u_{\mu}a_{\mu}^{2}}\mbox{Tr}U_{\mu}(r)
\label{eqn.3}
\end{equation}
Since the temporal lattice spacing in our simulations is very small, we 
adopt the following convention \cite{mor97,mor96,morn97} for the mean 
fields in 
tadpole 
improvement
\begin{equation} 
u_{t} \equiv 1, \hspace{1.0cm} u_{s} = \langle P_{ij}\rangle^{1/4}.
\label{eqn4}
\end{equation}                                                      
This prescription eliminates the need for  gauge fixing and the results 
yield values for $u_{s}$ which differ from those using Landau gauge 
by only a few percent.

\section{Renormalization of Anisotropy}
Following the procedure of Klassen \cite{klas98} and Shakespeare and 
Trottier \cite{shakes98},
we measure the static quark 
potential extracted from Wilson loops in the spatial and temporal 
directions. Accordingly on an anisotropic 
lattice there are  two potentials, $V_{xt}(R)$ and $V_{xy}(R)$. 
The two 
potentials differ by a factor of $\xi_{phys.}$ and by an additive constant, 
since 
the self-energy corrections to the static potential are different if the 
quark and anti-quark propagate along the temporal or a spatial direction. 
Thus $\xi_{phys.}$ can be determined by comparing the static quark 
potential 
computed from the logarithmic ratio of time-like Wilson loops 
$R(x,\tau)$, where
\begin{equation}
R(x,\tau) \equiv \frac{W_{xt}(x,\tau +1)}{W_{xt}(x,\tau )},
\label{eqn5}
\end{equation}
with the potential computed from that of the space-like Wilson loops 
$R(x,y)$, where
\begin{equation}
R(x,y) \equiv \frac{W_{xy}(x,y+1)}{W_{xt}(x,y)}.
\label{eqn6}
\end{equation}  
Asymptotically, for large $\tau$ and $y$, the ratios $R(x,\tau)$ and 
$R(x,y)$ approach
\begin{eqnarray}
R(x,\tau ) & = & Z_{x\tau}\mbox{e}^{-\tau V_{xt}}+(\mbox{excited state 
contr.})\\
R(x,y) & = & Z_{xy}\mbox{e}^{-yV_{xy}}+(\mbox{excited state
contr.}).
\label{eqn8}
\end{eqnarray}
To suppress the excited state contributions, a simple APE smearing 
technique \cite{alb87,teper86,ishi83} was used. 
In this technique an iterative smearing procedure is used to construct 
Wilson loop (and glueball) operators with a very high degree of overlap 
with the lowest-lying state. In our single-link smoothing procedure, we 
replace every space-like link variable by
\begin{equation}
U_{i}\rightarrow P\left[\alpha U_{i}+\frac{(1-\alpha 
)}{2}\sum_{s}U_{s}\right]
\label{eqn9}
\end{equation}
where the sum over `$s$' refers to the ``staples", or 3-link paths 
bracketing the given link on either side in the spatial plane, and P 
denotes a projection onto the group $U(1)$, achieved by renormalizing the 
magnitude to unity. We used a smearing parameter $\alpha =0.7$ and up to 
ten iterations of the smearing process. To reduce the statistical errors, 
the time-like Wilson loops were constructed from ``thermally averaged"
time-like links \cite{teper86,ishi83}.

The links making up the space-like 
and time-like Wilson loops  are smeared by the same amount so that the 
ratios $R(x,\tau)$ and $R(x,y)$ have the same excited-state contribution. 
Similarly, finite-volume corrections to the $R(x,\tau )$ and $R(x,y)$ are 
the same if the temporal and spatial extents are equal in physical units, 
i.e. $N_{s}=\xi_{phys.} N_{t}$ in lattice units. These statements are 
expected to 
hold only for large $x$, $y$ and $\tau $; otherwise there can be large 
$O(a^{2}_{s},a^{2}_{t})$ lattice errors.

The physical anisotropy is determined from the ratio of the potentials
$V_{xt}(R)$ and $V_{xy}(R)$ estimated from $R_{xt}$ and $R_{xy}$ respectively.
 The unphysical constant in the potentials is 
removed by subtraction of the simulation results at two different 
radii
\begin{equation}
\xi_{phys.} = 
\frac{V_{xt}(R_{2})-V_{xt}(R_{1})}{V_{xy}(R_{2})-V_{xy}(R_{1})}.
\end{equation}
The measured renormalization of the anisotropy,  
$\eta$ is then determined from
\begin{equation}
\eta \equiv \frac{\xi_{phys.}}{\xi_{0}}.
\end{equation}

\section{Simulation and Results}
Simulations were performed on four lattices of $N_{s}^{2}\times N_{t}$ 
sites, with $N_{s}=16$ and $N_{t}$ ranging from 32 to 48  with mean-link 
improvement, and 
four lattices with the Wilson action. Configurations were generated 
by using the Metropolis algorithm. The details of the algorithm are 
discussed elsewhere \cite{loan02}. 50000 sweeps were performed for thermalization of the 
configurations 
and self-consistent determination of the tadpole factors. Configurations are stored every 
250 sweeps thereafter.  Ensembles of about 1000 
configurations were used to measure the static quark potential, while 
1,400 configurations, at coupling values from $\beta = 1.0$ to 2.5, were 
generated for the 
glueball mass. We fixed $\xi_{0} =16/N_{t}$ in the first pass, so that the 
lattice size remains fixed at $16a_{s}$ in all directions. The 
simulation parameters of the lattices analyzed here are given in Table 
 \ref{tadpara}. 

\begin{table}[!h]
\caption{
\label{tadpara}
Simulation parameters for the lattices analyzed for
renormalization of the anisotropy. The bare anisotropies $\xi_{0}$ and the 
means fields $u_{t}$ and $u_{s}$ for tadpole improvement are shown.}
\begin{ruledtabular}
\begin{tabular}{cccccc}
Action & $\xi_{0}=\Delta\tau$ & $\beta$ & $u_{t}$ & $ u_{s}$ & Volume \\
\hline
Tadpole  & 0.50 & 1.306  & 1. & 0.924 & $16^{2}\times 32$\\
Improved &      & 1.443  & 1. & 0.931 & \\
         &      & 1.592  & 1. & 0.937 & \\
         &      & 1.892  & 1. & 0.946 & \\
         & 0.40 & 1.302  & 1. & 0.921 & $16^{2}\times 40$\\
         &      & 1.436  & 1. & 0.927 & \\
         &      & 1.584  & 1. & 0.932 & \\
         &      & 1.88   & 1. & 0.940 & \\
         & 0.33 & 1.2931 & 1. & 0.915 & $16^{2}\times 48$\\
         &      &1.426   & 1. & 0.920 & \\
         &      &1.567   & 1. & 0.922 & \\
         &      &1.858   & 1. & 0.929 & \\
Wilson   & 0.50 & 1.4142 &    &       & $16^{2}\times 32$\\
         & 0.40 &        &    &       & $16^{2}\times 40$\\
         & 0.33 &        &    &       & $16^{2}\times 48$\\  
\end{tabular}
\end{ruledtabular}
\end{table}

\begin{figure}[!h]
\scalebox{0.45}{\includegraphics{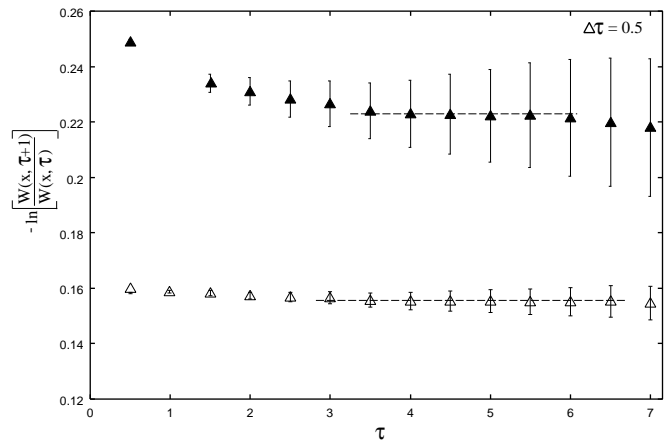}}
\caption{
\label{fig1}
Logarithmic ratio of the time-like Wilson loops as a function of 
$\tau$ for fixed $x$ at $\beta =1.309$ and $\Dtau =0.5$. The solid triangles
correspond to $x=4$ and open triangles show $x=2$ values.}
\end{figure}

After measuring the Wilson loops at fixed values of $\beta$, we compute 
the ratios $R_{xt}$ and $R_{xy}$. We find that the individual ratios reach 
their plateaus for $\tau \geq 3$ and $y\geq 3$ for fixed $x$ as shown in 
Figures \ref{fig1} and \ref{fig2}. These ratios are
are expected to be independent of $\tau$ and $y$ for $\tau$, 
$y$ $\geq 3$ respectively.
The estimates of the potentials $V_{xt}(R)$ and $V_{xy}(R)$ can  now be 
found from these ratios.

\begin{figure}[!h]
\scalebox{0.45}{\includegraphics{Fig2}}
\caption{
\label{fig2}
Logarithmic ratio of the spatial Wilson loops as a function of 
$y$ for fixed $x$ at $\beta =1.309$ and $\Dtau =0.5$. The solid triangles
correspond to $x=4$ and open triangles show $x=2$ values.}
\end{figure}

\begin{figure}[!h]
\scalebox{0.45}{\includegraphics{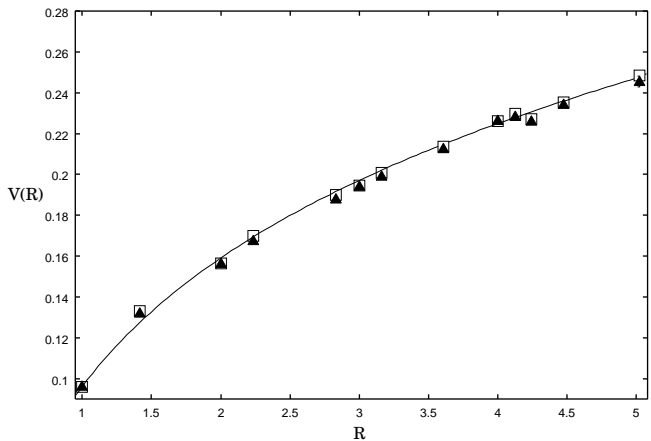}}
\caption{
\label{fig3}
Static quark potentials computed from $R_{xt}$ (solid triangles) and
$R_{xy}$ (open squares) as a function of separation $R$ at $\beta =1.306$ and
$\Dtau =0.5$ using mean field improved gauge action.
The potential obtained from $R_{xt}$ has been rescaled 
by the input anisotropy. The solid line is a fit  to 
the form $V(R) =a+\sigma R+c\mbox{ln}(R)$ to the temporal Wilson loops.}
 \end{figure}

Figure \ref{fig3} shows a  graph of the static quark potentials, computed 
from spatial and temporal Wilson loops,  as a
function of radius $R$ at $\beta = 1.306$ and $\Delta \tau = 0.5$. The 
potential in the lattice units obtained from the ratio of the time-like 
Wilson loops has been rescaled by the input anisotropy.
To extract the string tension, the time-like potential  is well 
fitted by a form
\begin{equation}
V_{xt}(R) = a +b\ln R+\sigma R,
\label{eqn27}
\end{equation}                                
including a logarithmic Coulomb term as expected for classical
QED in (2+1) dimensions which dominates the behaviour at small distances,
and
a linear term as predicted by Polyakov \cite{pol78} and G{\" o}pfert and
Mack \cite{gof82} dominating the behaviour at large distances and showing 
a clear evidence of the
linear confining behaviour at large distances.

\begin{figure}[!h]
\scalebox{0.45}{\includegraphics{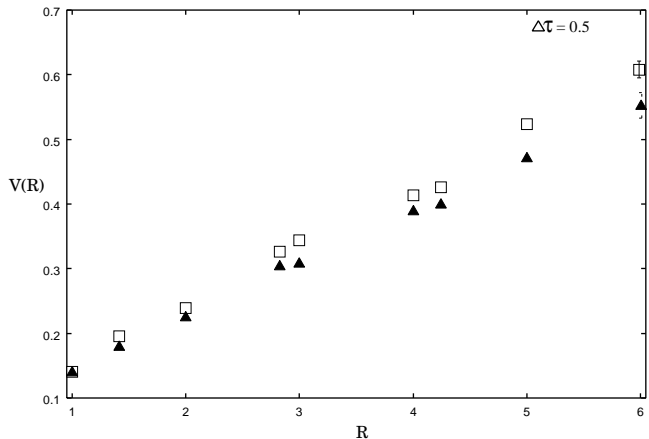}}
\caption{
\label{fig4}
Static quark potentials computed from $R_{xt}$ (solid triangles) and
$R_{xy}$ (open squares)  at $\beta =1.306$ and
$\Dtau =0.5$ for standard Wilson action. }
\end{figure}

We measured each anisotropy twice, using two different radii $R_{1}$ for 
subtraction, with fixed $R_{2}$. Setting $R_{2}=4$, we computed the 
anisotropy with $R_{1} = 2$ and $\sqrt{2}$. The two determinations  
of anisotropy, shown in Table \ref{tadres}, are in excellent agreement.
The numerical values of the renormalization of the anisotropy 
parameter $\eta$ appears 
to be equal to unity even at large $\beta$. 
It is seen that for with mean-field improvement  the input 
anisotropy
is renormalized  by few percent over the range of lattices analyzed 
here, whereas  the measured value of anisotropy is about $15-20\%$ 
lower than the bare anisotropy with the standard Wilson action.
This can be seen from Figure \ref{fig4} where the renormalization of 
the anisotropy is plainly visible as a difference in slope of the 
potentials 
computed from $R_{xt}$ and $R_{xy}$. 

\begin{table}[!h]
\caption{
\label{tadres}
Measured anisotropy $\xi_{phys.}$ compared to the bare anisotropy
$\xi_{0}$ for two actions.}
\begin{ruledtabular}
\begin{tabular}{ccccccc}
Action &  $\xi_{0}$ & $\beta$ & \multicolumn{2}{c}{$\xi_{phys.}$} &
\multicolumn{2}{c}{$\eta =\xi_{phys.}/\xi_{0}$}  \\
& & & $R_{1}=2$ & $R_{1}=\sqrt{2}$ & $R_{1}=2$ & $R_{1}=\sqrt{2}$  \\ \hline
Tadpole  & 0.50 & 1.306  & 0.500(1)  & 0.500(2) &1.00(1) &1.00(4) \\
Improved &      & 1.443  & 0.501(2)  & 0.503(2) &1.00(4) &1.00(4)  \\
         &      & 1.592  & 0.50(2)   & 0.50(3)  &1.00(4) &1.00(6) \\
         &      & 1.892  & 0.50(5)   & 0.50(4)  &1.0(1)  &1.00(8)\\
         & 0.40 & 1.302  & 0.402(2)  & 0.401(2) &1.00(4) &1.00(5) \\
         &      & 1.436  & 0.402(1)  & 0.403(8) &1.00(2) &1.00(1) \\
         &      & 1.584  & 0.40(3)   & 0.40(2)  &1.00(6) &1.00(4) \\
         &      & 1.88   & 0.40(1)   & 0.40(1)  &1.00(7) &1.00(2) \\
         & 0.33 & 1.293  & 0.335(3)  & 0.34(1)  &1.00(8) &1.03(3) \\
         &      & 1.426  & 0.33(2)   & 0.33(1)  &1.00(6) &1.00(3)  \\
         &      & 1.567  & 0.33(2)   & 0.33(1)  &1.00(6) &1.00(3) \\    
         &      & 1.858  & 0.33(3)   & 0.33(1)  &1.00(6) &1.00(3) \\
Wilson   & 0.50 & 1.414  & 0.482(4)  & 0.471(3) &0.964(8)&0.942(6) \\
         & 0.40 & 1.414  & 0.336(4)  & 0.338(3) &0.84(1) &0.845(7) \\
         & 0.33 & 1.414  & 0.279(5)  & 0.281(2) &0.83(1) &0.843(6)\\
\end{tabular}
\end{ruledtabular}
\end{table}

Glueball correlation functions $C(\tau )$ were also calculated 
\begin{equation}
C(\tau )=\sum_{\tau_{0}} \langle0\mid \bar{\Phi}_{i}(\tau 
+\tau_{0})\bar{\Phi}_{i}(\tau_{0})\mid 0\rangle
\end{equation}
where $\bar{\Phi}_{i}(\tau )$ is the optimized glueball operator 
found by a variational technique, following Morningstar and Peardon 
\cite{morn97} and Teper \cite{tep99}, from a  linear combination of the 
basic operators $\phi_{i}$, 
\begin{equation}
\Phi (\tau ) =\sum_{\alpha}v_{i\alpha}\phi_{i\alpha}(\tau )
\end{equation}
where the index $\alpha$ runs over the rectangular Wilson loops with 
dimensions $l_{x}=[n-1,n+1]$, $l_{y}=[n-1,n+1]$ and smearing 
$n_{s}=[m-1,m+1]$, making 27 operators in all.  

The optimized correlation function was fitted with the simple form
\begin{equation}
C_{i}=c_{0}+c_{1}\cosh m_{i}(T/2 -\tau )
\end{equation}
to determine the glueball mass estimates.

\begin{figure}[!h]
\scalebox{0.45}{\includegraphics{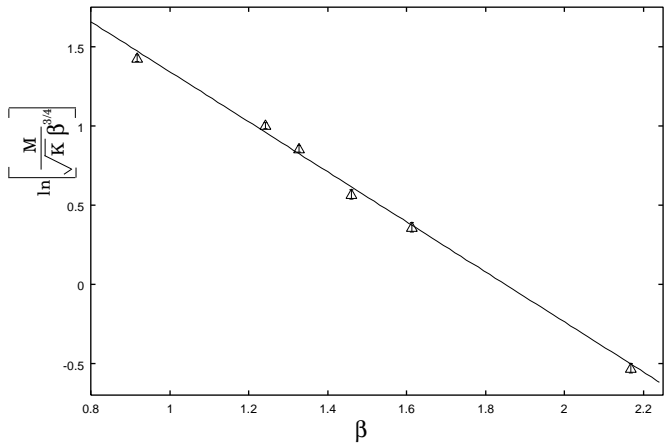}}
\caption{
\label{fig5}
Logarithmic  dimensionless ratio $M/\sqrt{K}\beta^{3/4}$
as a function of $\beta$. The solid curve is the fit to the data for 
$0.916\leq \beta\leq 2.12$.}
 \end{figure}

The results for the symmetric and the anti-symmetric glueball masses 
over the square root of the string tension are
shown in Table \ref{tadmass}, along with the mean plaquette values at
different $\beta$ at $\Dtau =1.0$.  
Figure \ref{fig5}   shows the behaviour of the logarithm of the antisymmetric 
mass 
gap over the square root of the string tension as a 
function of $\beta$. It can be seen that that the ratio  scales  
exponentially to zero in the weak-coupling limit as should be in 
three-dimensional confining theories. 
The solid line is a fit to the data over the range
$0.916\leq \beta \leq 2.12$. The slope of the data matches the 
predicted form \cite{gof82}, however, the intercept of the scaling curve is large by a 
factor of 2 (our previous estimates of constant coefficient 
for standard Wilson action are large by a factor of 5.2 \protect\cite{loan02}).
It would be interesting 
to test the sensitivity of the slope and intercept of the scaling curve by 
including the radiative corrections.

\begin{figure}[!h]
\scalebox{0.45}{\includegraphics{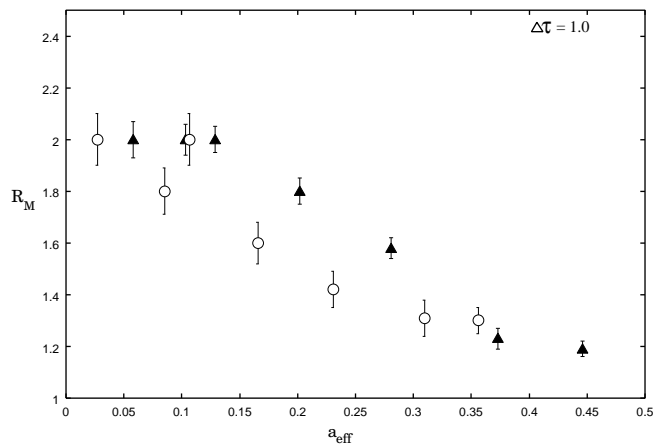}}
\caption{
\label{fig6}
Mass ratio $M_{R}$ against the effective lattice spacing $a_{eff}$ at
$\Dtau =1.0$. The solid triangles show the estimates for mean field improved
gauge action and open circles show the estimates of standard Wilson action.}
\end{figure}
A plot of mass ratio 
against the effective lattice spacing 
$a_{eff}$ \cite{loan02} is shown in Figure \ref{fig6}.
At weak coupling, the theory is expected to approach a theory of free 
bosons \cite{gof82} so that symmetric state will be composed of two 
$0^{--}$ bosons and the mass ration should approach two.  The mass
ratio of the lowest glueball states scale against the effective lattice
spacing towards a value close to 2.0, as expected for a theory of free
scalar bosons. We note that mass ratio exhibits  scaling behaviour, 
even with the Wilson action \cite{loan02}, however, in contrast with the 
Wilson action, a significant 
reduction in the errors with tadpole improved action is apparent in the 
mass gap and the mass ratio.

\begin{table}[!h]
\caption{
\label{tadmass}
Monte Carlo estimates for the mean plaquette $\langle P\rangle$, the effective
lattice spacing, the
symmetric and antisymmetric glueball masses $M_{0^{++}}$, $M_{0^{--}}$,
for tadpole improved Wilson action at $\Delta\tau = 1.0$}
\begin{ruledtabular}
\begin{tabular}{ccccccc}
Action & $\beta$ & $\langle P\rangle$ &$a_{eff}$ & $M_{0^{++}}/\sqrt{K}$ & 
$M_{0^{--}}/\sqrt{K}$
& $R_{M}$  \\ \hline 
Tadpole & 0.917 & 0.678 & 0.864 & 4.0(1) & 3.9(1) & 1.03(3) \\
improved &1.243 & 0.719 & 0.446 & 3.83(9) & 3.2(1) & 1.19(3)\\
       & 1.327 & 0.776 & 0.373 & 3.5(2) & 2.9(1) & 1.23(4)\\
       & 1.460 & 0.797 & 0.281 & 3.7(1) & 2.34(7) & 1.58(4)\\
       & 1.613 & 0.816 & 0.202 & 3.6(1) & 2.05(2) & 1.80(5)\\
       & 1.816 & 0.836 & 0.129 & 2.05(4) & 0.98(3) & 2.00(5)\\
       & 1.917 & 0.844 & 0.103 & 2.0(1) & 0.99(5) & 2.00(6) \\
       & 2.168 & 0.861 & 0.058 & 2.19(6) & 1.05(2) & 2.08(7) \\   
       & 2.417 & 0.874 &       &         &         &  \\
       &2.669  & 0.885 &       &       &   & \\
       &2.917  & 0.895 &       &         & \\ 
Wilson & 1.0   & 0.475 &0.733 &3.27(3) & 3.1(1)  & 1.03(6) \\
       & 1.35  & 0.629 &0.356 &3.6(1)  & 2.81(9) & 1.30(5)\\
       & 1.41  & 0.656 &0.310 &3.8(1)  & 2.95(8) & 1.31(7)\\
       & 1.55  & 0.704 &0.231 &4.0(1)  & 2.81(6) & 1.42(7)\\
       & 1.70  & 0.748 &0.166 &4.03(8) & 2.51(2) & 1.60(8)\\
       & 1.90  & 0.790 &0.107 &3.9(1)  & 1.88(3) & 2.0(1)\\
       & 2.0   & 0.806 &0.085 &3.53(4) & 1.96(4) & 1.80(9)\\
       & 2.5   & 0.854 &0.027 &3.1(2)  & 1.50(8) & 2.0(1)\\ 
\end{tabular}
\end{ruledtabular}
\end{table}

\section{Summary and Outlook}
Mean field improved $U(1)$ lattice gauge theory in (2+1) dimensions was 
applied to  calculations of the static quark potential, the renormalized 
lattice anisotropy and the scalar glueball masses. 
We analyzed the mean-link improved action on isotropic and anisotropic 
lattices and comparisons were made with the simulations of the Wilson 
action. By comparing the static quark potentials computed from space-like 
and time-like Wilson loops, we determined the physical anisotropy of the 
tadpole improved Wilson action. We found that with mean-link improved 
Wilson action, the bare anisotropy is renormalized by less than 
a few percent, in contrast with the standard Wilson action, where the 
measured value of anisotropy is found to be about $15-20\%$ lower than 
bare anisotropy on the lattices analyzed here. We found that tadpole 
improvement significantly reduces discretization errors in the static 
quark potential and the glueball masses. The mass ratio of the two lowest 
glueball states scales against the effective lattice spacing towards a 
value close to 2.0, as expected for a theory of free scalar bosons.

We intend to extend PIMC  techniques to Symanzik  improved 
$U(1)$ lattice gauge theory. The intention is  to study the effects of 
improvement on the scaling slope, the constant coefficients and
scaling behaviour observed in the weak-coupling regime of the theory.
We also plan the study the one-loop correction to the anisotropy factor for
Symanzik improved $U(1)$ gauge action in three dimensions. We shall report
on this work in the near future.

\begin{acknowledgments}
This work was supported by the Australian
Research Council. We are  grateful for access to the
computing facilities of the Australian Centre for Advanced Computing and   
Communications (ac3) and the Australian Partnership for Advanced
Computing (APAC).
\end{acknowledgments}

\end{document}